\documentclass[9pt,twocolumn]{article}

\usepackage[utf8]{inputenc}
\usepackage[T1]{fontenc}
\usepackage{times}
\usepackage[a4paper,margin=1in]{geometry}

\usepackage{cite}
\usepackage{amsmath,amssymb,amsfonts}
\usepackage{algorithmic}
\usepackage{graphicx}
\usepackage{textcomp}
\usepackage{xcolor}
\usepackage{booktabs}
\usepackage{multirow}
\usepackage{colortbl}
\usepackage{float}
\usepackage{wrapfig}
\usepackage{caption}
\usepackage{subcaption}
\usepackage{siunitx}
\usepackage{hyperref}
\usepackage{tabularx}

\graphicspath{{images/}{figures/}{plots/}}

\usepackage{soul}
\usepackage{comment}

\soulregister\ref7
\soulregister\pageref7
\soulregister\eqref7
\soulregister\cite7
\soulregister\qty7
\soulregister\SI7

\newcommand{\review}[1]{{\hl{#1}}}

 \renewcommand{\review}[1]{#1}

\usepackage{enumitem}
\usepackage{pifont}

\newlist{todolist}{itemize}{2}
\setlist[todolist]{label=$\square$}

\definecolor{abstractbg}{rgb}{1,0.969,0.914}
\definecolor{darkgreen}{RGB}{0,120,0}

\def\BibTeX{%
  {\rm B\kern-.05em{\sc i\kern-.025em b}\kern-.08em
  T\kern-.1667em\lower.7ex\hbox{E}\kern-.125emX}%
}

\begin{document}

\title{WULPUS PRO: Multi-mode Ultra-Low-Power Wearable Ultrasound and Array Imaging with CMUT Support}

\author{%
Sergei Vostrikov$^{1,2}$,
Federico Villani$^{1}$,
Cedric Hirschi$^{1}$,
Jinhao Lu$^{3}$,
Jonas Welsch$^{3}$,\\
Martin Angerer$^{3}$,
Edmond Cretu$^{3}$,
Robert Rohling$^{3}$,
Andrea Cossettini$^{1}$,
and Luca Benini$^{1,4}$\\[1ex]
\small $^{1}$Integrated Systems Laboratory, ETH Z{\"u}rich, Switzerland, \small $^{2}$R\&D Department, Rheonics, Winterthur, Switzerland\\
\small $^{3}$Department of Electrical and Computer Engineering, The University of British Columbia, Vancouver, Canada\\
\small $^{4}$Department of Electrical, Electronic, and Information Engineering (DEI), University of Bologna, Italy\\
\small Contact: Sergei Vostrikov, sergei.vostrikov@rheonics.com
}

\date{}

\maketitle

\begin{abstract}

Wearable ultrasound has emerged as a key technology for continuous monitoring of physiological processes such as muscle dynamics, bladder volume, and cardiovascular activity. Existing fully-wearable ultra-low-power platforms are limited to shallow, low-channel A-mode sensing, while large-channel count ($>$8) multi-mode systems are too large and power-hungry for true wearability. To bridge this gap, we present WULPUS PRO, a novel runtime-programmable wearable ultrasound acquisition platform with an ultra-compact size of $39\times21\times6\,\mathrm{mm}$ and low weight (5\,g) that integrates 30\,V excitation, 16 time-multiplexed channels, a low-noise receive front-end with up to 70\,dB gain ({14}\,MHz amplification chain bandwidth),  time-gain compensation, and SNR of {45\,dB}. This enables deep-tissue echo acquisition using transducers with central frequency up to 2.2\,MHz in RF-sampling mode and up to 8\,MHz in envelope-detection mode. We demonstrate for the first time B-mode imaging in a 16-channel ultra-low-power wearable, achieving \review{sub-millimeter axial and millimeter-scale lateral resolution in phantom experiments} while operating within only 40 mW power budget at 50\,Hz \review{Pulse Repetition Frequency} (PRF) and under 60 mW at 300 Hz \review{PRF}. Furthermore, WULPUS PRO demonstrates support, for the first time, of both piezoelectric and Capacitive Micromachined Ultrasonic transducers (CMUTs), enabling tight integration with emerging skin-conformal polymer-based CMUT arrays. \review{WULPUS PRO is designed as a host-agnostic acquisition front-end, exposing standard data and power interfaces for integration with different wearable host platforms (e.g., BLE- or WiFi-based)}. 
We demonstrate wireless data transmission with external Bluetooth Low Energy and Wi-Fi modules, \review{and we project} multi-day BLE operation (1-2 days at 50\,Hz PRF) and multi-hour Wi-Fi streaming ($>3$ h at 300\,Hz PRF) \review{for} a 300\,mAh \review{(6.4\,g)} Li-Po cell. Thus, WULPUS PRO establishes a new class of wearable ultrasound acquisition platforms for continuous, fully programmable, B-mode-enabled, ultra-low-power and low-noise sensing.

\end{abstract}

\begin{figure}[!t] 
    \centering 
    \includegraphics[width=0.5\textwidth]{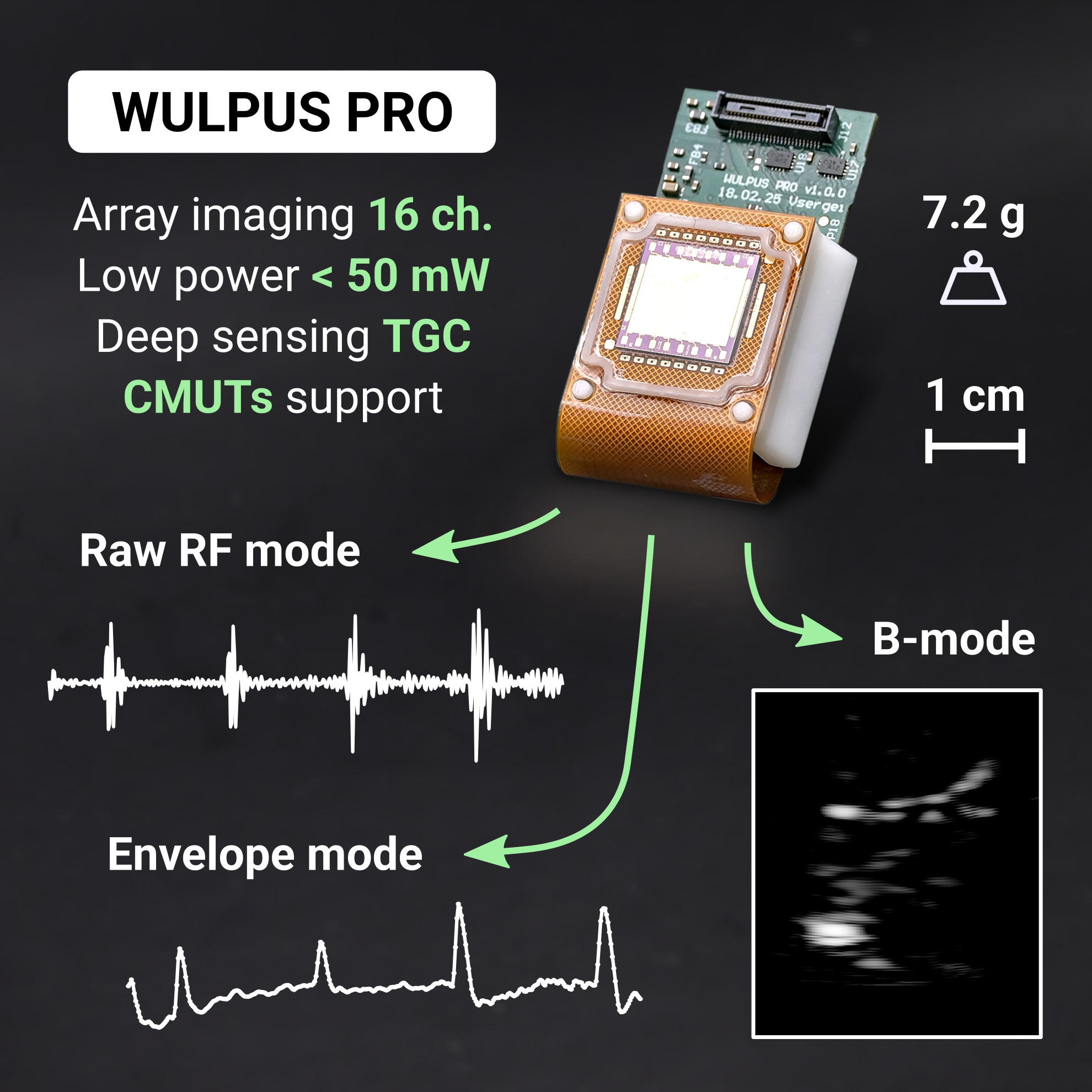} 
    \end{figure}

\begin{samepage}
    \par
\end{samepage}

This work was supported by the ETH Research Grant ETH-C-01-21-2,
Project ListenToLight.

\section{Introduction}
\label{sect:intro}

Wearable ultrasound (US) is emerging as a powerful tool for continuous, unobtrusive body monitoring \cite{huang2023emerging}. Handheld US systems have demonstrated increasing impact in point-of-care diagnostics \cite{baribeau2020handheld}, but they remain bulky, power-hungry, and unsuitable for long-term continuous body-worn deployment. Fully-wearable systems, instead, enable continuous access to deep-tissue information with minimal impact on the user, enabling applications such as hand and finger movement tracking \cite{yang_wearable_2021, lu2022wearable, zhang2024live, vostrikov2024unsupervised}, cardiovascular monitoring \cite{kenny2021novel, frey2022wulpus, lin2024fully,zhou2025clinical,peng2025wearable}, and monitoring the urinary bladder \cite{van2019sens, vostrikov2024tinyprobe,cai2025real}.

Despite this growing number of use cases, current solutions present significant limitations. Commercial devices demonstrated the clinical viability of wearable US \cite{flopatch,dfree}, but they remain single-purpose, closed systems, lacking programmability, without providing access to raw data. This typically prevents their use for new research developments.

\definecolor{subsectioncolor}{RGB}{0,0,0}

\begin{table*}[!t]
\arrayrulecolor{subsectioncolor}
\setlength{\arrayrulewidth}{1pt}
\renewcommand{\arraystretch}{1.15}
\setlength{\tabcolsep}{4pt}

{\sffamily
\begin{tabularx}{\textwidth}{@{}p{2em}X@{}}
\hline

\rowcolor{abstractbg}
\multicolumn{2}{@{}l@{}}{%
    \color{subsectioncolor}\bfseries\itshape Highlights\strut
}\\[0.5em]

\rowcolor{abstractbg}
$\bullet$ &
We present WULPUS PRO, the first ultra-low-power, fully wearable ultrasound
acquisition platform enabling 16-channel B-mode imaging, operating at under
40\,mW (50\,Hz PRF) and below 60\,mW (300\,Hz PRF) in a compact
$39 \times 21 \times 6$\,mm, 5\,g form factor.
\\[1em]

\rowcolor{abstractbg}
$\bullet$ &
WULPUS PRO integrates runtime-programmable 30\,V TX excitation and low-noise RX chain featuring time-gain compensation, up to 70\,dB gain, {14}\,MHz analog bandwidth, and optional envelope detection.
\\[0.6em]

\rowcolor{abstractbg}
$\bullet$ &
The platform uniquely supports PZT and CMUT transducers, enabling modular
multi-wireless operation with 1--2 days of BLE use (50\,Hz PRF) or over
3\,h of Wi-Fi streaming (300\,Hz PRF) from a 300\,mAh battery.
\\[0.6em]

\hline
\end{tabularx}
}

\setlength{\arrayrulewidth}{0.4pt}
\arrayrulecolor{black}
\end{table*}

While an increasing number of wireless, wearable research platforms enable access to raw data, they exhibit severe design trade-offs. On one hand, higher performance systems consume hundreds of milliwatts \cite{lin2024fully,vostrikov2024tinyprobe}, thus \review{limiting the continuous operation to only a few hours}. On other hand, lower-end platforms such as the open-source WULPUS probe achieve multi-day battery lifetime, yet at the price of reduced channel count and supported imaging modalities \cite{frey2022wulpus}.
In parallel, advances in emerging transducer technologies \cite{vanneer2024flexibleus,zhou2023continuous,keller2023fully,xue2023flexible} highlight the need for open, reconfigurable wearable platforms that can support a variety of transducers.

To address these challenges and building upon our earlier WULPUS design, this work introduces WULPUS PRO, an ultra-low-power wearable platform with expanded functionality, programmability, and transducer support
. This paper proposes the following main contributions:
\begin{itemize}
    \item Design of a new 16-channel wearable US architecture ($2\times$ more channels than WULPUS) supporting both piezoelectric and 
    Capacitive Micromachined Ultrasonic Transducer (CMUT) transducers through reconfigurable biasing options and programmable amplifier stages;
    \item Redesigned high-voltage pulsing stage for operation up to 30\,V unipolar ($2\times$ higher than WULPUS), coupled with a new low-noise signal amplification chain that introduces a high dynamic range Variable Gain Amplfier (VGA), Time Gain Compensation (TGC), and optional analog envelope detection to support transducers up to 8\,MHz, guaranteeing a mid-band signal-to-noise ratio (SNR) as high as 32\,dB;
    \item Expanded modality support, enabling A-mode, M-mode, B-mode (synthetic aperture), and envelope detection;
    \item \review{Host-agnostic front-end exposing standard data and power interfaces, while locally generating the high-voltage domains, enabling integration with different wearable hosts over BLE or WiFi;}
    \item Ultra-low-power operation, consuming less than 50\,mW at 50\,Hz PRF and less than 60\,mW at frame rates as high as 300\,Hz, thereby enabling multi-day operation when operated via BLE and multi-hour use via WiFi on a compact 300\,mAh LiPo battery;
    \item Open-source release with a permissive license of all hardware and software design files: \url{https://github.com/pulp-bio/wulpus-pro}
\end{itemize}

The paper is structured as follows. Sect.~\ref{sect:related} presents related works. Sect.~\ref{sect:methods} focuses on WULPUS PRO design. Sect.~\ref{sect:electronics_characterization} presents its hardware characterization, in terms of power consumption, bandwidth, and SNR. Sect.~\ref{sect:functional_evaluation} validates the functionality of WULPUS PRO across different operating modes. Sect.~\ref{sect:integ_polycmuts} discusses integration with polymer-based CMUTs (polyCMUTs).
Finally, Sect. \ref{sect:soa_comparison}-\ref{sect:discussion} compare WULPUS PRO to state-of-the-art (SoA), discussing the proposed advancements and current limitations, and concluding the paper.
\section{Related Works}
\label{sect:related}

In recent years, commercial wearable US systems have started to appear on the market, primarily as single-purpose clinical devices rather than multi-use sensing platforms.
Representative examples include the FloPatch \cite{flopatch}, a single-mode platform for real-time assessment of carotid arterial pulses using Continuous Wave (CW) doppler; DFree \cite{dfree} together with SENS-U \cite{van2019sens}, two compact bladder monitoring devices to estimate bladder volume and manage incontinence. These systems have found broad clinical deployment, demonstrating the feasibility of using wearable US patches in routine care.
Several additional products are also progressing toward the market, such as the Slanj from Novosound \cite{novosound}, used for cuffless blood pressure estimation, as well as the Cynova \cite{cynova} and Sonus \cite{sonus} cardiac patches. 
However, all existing solutions share key limitations, such as single operation mode, lack of programmability and configurability, and lack of access to raw data, thereby limiting new algorithmic developments by the US community.

Beyond commercial devices, the interest of the research community in wearable US has been steadily increasing, leading to a diverse set of novel transducers and systems.
Notable examples include the works of Steinberg et al. \cite{steinberg2022continuous} targeting continuous artery monitoring, Zhou et al. \cite{zhou2025clinical} with a solution for non-invasive monitoring of blood pressure,  
Wu et al. \cite{hu2023wearable} for cardiac imaging, Xue et al. \cite{xue2023development} for muscular activity, and Qu et al. \cite{qu2024continuously} for muscle fatigue.
While these works demonstrate remarkable skin conformity and imaging quality, they primarily focus on the transducer technology, relying on bench-top electronics for signal acquisition.

\review{Only a limited number of complete wearable US systems have been reported. For such systems, battery lifetime is a key practical constraint, largely determined by the average power consumption of the electronics. When comparing these platforms, reported power values should be interpreted with care, as the literature does not always clearly distinguish between instantaneous active power and average power over a duty-cycled operating profile. In the following review of complete wearable US solutions, we report the available total system power values when specified and treat them as estimates of average power consumption under the operating conditions described in each work.}

Zou et al. \cite{zou2024fully} is a recent example of a wearable US therapeutic device. The authors present a wearable solution for continuous sonodynamic therapy, with a patch that focuses US on lesions to treat tumors. 

Toymus et al. \cite{toymus2024integrated} presented  an integrated wearable US bladder volume monitoring device, incorporating  flexible and air-backed transducers with compact control electronics for wireless data transmission via bluetooth low-energy (BLE) to a mobile phone. The platform is based on the STM32WB55RGV6 microcontroller, features 4 channels (multiplexed), and consumes approx. 150-200\,mA, offering a lifetime of approx. 10 hours when operated with a 2000\,mAh battery.

The UsoP platform \cite{lin2024fully} offers increased performance by including a larger number of channels (32-channels for Transmit (TX), 1 channel at a time for Receive (RX)), multi-modality (A/B/M-modes), an increased sampling rate of 12\,Msps, and a wireless WiFi link (3.4\,Mbps). USop features a compact form factor (3.5$\times$30$\times$83\,mm) and a total power consumption of 614\,mW.

Further advancements have been enabled by TinyProbe \cite{vostrikov2024tinyprobe}, thanks to its multi-channel nature (32-channels individually accessible for TX and RX), higher sampling rate (30\,Msps), high-PRF (1400\,Hz) Doppler capabilities, and increased wireless bandwidth (35.7\,Mbps \cite{hirschi2025high}) with a state-of-the-art transmission efficiency of 73.5\,Mbps/W and a power consumption of 100 - 1240\,mW \cite{vostrikov2024tinyprobe}.

\review{These higher-performance platforms therefore provide broader functionality, but at the cost of increased system power.} In this context, WULPUS \cite{frey2022wulpus} appeared as the first open, programmable ultra-low-power wearable probe, offering A-mode capabilities (on 8-channels, time multiplexed) with BLE-based raw data access in less than 25 mW, thereby enabling multi-day continuous monitoring. Thanks to the open-source approach \cite{wulpus_repo}, WULPUS has been extensively used for a large number of applications across research laboratories. Another recent example is PuLsE system \cite{giordano2025pulse}, optimized for ultra-low-power (5.8\,mW) single-channel artery monitoring in envelope mode, achieving continuous operation for over seven days on a smartwatch battery.

Overall, the current landscape shows promising trends toward wearable US, but a clear gap appears: existing open, programmable systems either offer ultra-low power with limited functionality (e.g., WULPUS, PuLsE), or include broader capabilities at the cost of higher power consumption (e.g., USoP, TinyProbe). Additionally, they typically support only a single class of transducers, and therefore cannot accommodate the growing diversity of emerging wearable transducers.
This paper addresses these challenges by presenting WULPUS PRO, the next-generation of our open-source, programmable, ultra-low-power wearable US platform, providing increased channel-count, extended voltage and frequency range, inclusion of TGC capabilities, additional configurability and modality support, and compatibility with multiple transducer technologies. 

\section{System Design}
\label{sect:methods}

\subsection{System Architecture}

\begin{figure*}[tb]
  \centering
  \includegraphics[width=1\textwidth, clip, trim={0 0 0 0 cm}]{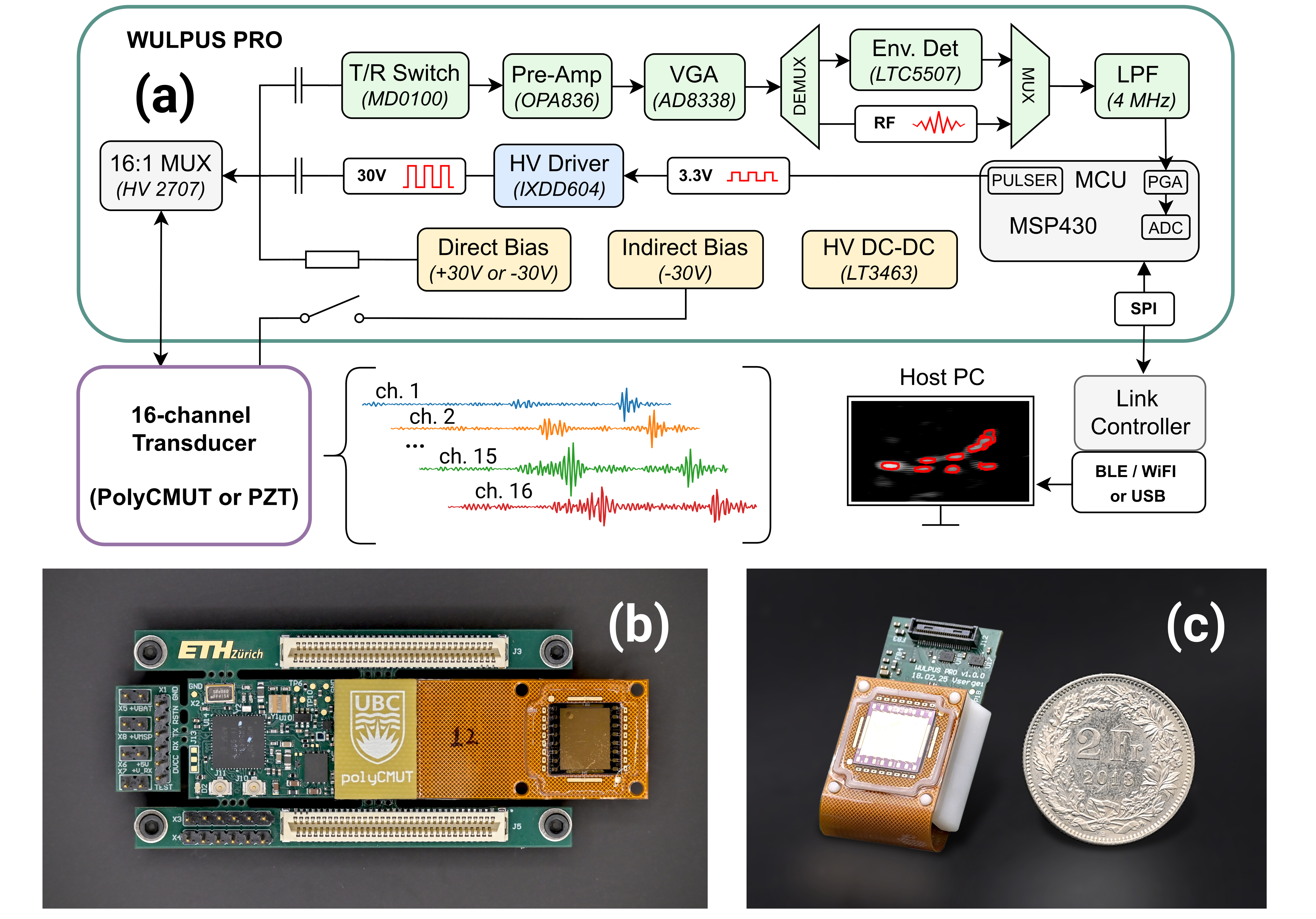}
  \caption{(a) WULPUS PRO system block diagram.
(b) WULPUS PRO evaluation board with a polyCMUT transducer attached. The board features breakaway sections that mirror programming, communication interfaces, and power supply pins for easy debugging.
(c) WULPUS PRO core module with a polyCMUT transducer shown next to a 2 Swiss franc coin for size reference.}
  \label{fig:system_diagram}
\end{figure*}

The proposed system builds on the MSP430FR5043-based ultrasound platform presented in the original WULPUS design~\cite{frey2022wulpus} and addresses its critical limitations by supporting deeper sensing, broader transducer range and higher frame rates. These improvements are realized through a redesigned transmit and receive paths, featuring a doubled excitation amplitude, compatibility with larger-aperture transducers (up to 16 channels), support for higher central frequencies (up to 8\,MHz) in envelope mode, and the integration of a time-gain compensation. In addition, WULPUS PRO introduces a modular architecture that decouples the ultrasound front end from the data transmission link, thereby enabling flexible connectivity options: Wi-Fi for fast frame rates, BLE for ultra-low-power battery operation, or wired link for robust long-term data acquisition.

The overall system architecture is illustrated in Fig.~\ref{fig:system_diagram} (a). During data acquisition, the MSP430 generates excitation pulses that are amplified by the IXDD604 driver (Littelfuse, Chicago, USA) and routed through the HV2707 high-voltage multiplexer (Microchip Technology, Chandler, USA) to the transducer array. After transmission, the IXDD604 is switched to a high-impedance state, and the multiplexer is reconfigured for receive, allowing echo signals to propagate through the MD0100 TX/RX switch (Microchip Technology, Chandler, USA) and reach a variable-gain amplifier (VGA) AD8338 (Analog Devices, Wilmington, USA), implementing time-gain compensation (TGC). Next, the received signal can optionally undergo envelope detection via the LTC5507 before being low-pass filtered and digitized by the MSP430’s 8 Msps sigma-delta analog-to-digital converter (ADC). After sampling, ultrasound data is transmitted via an 8-MHz SPI link to the host PC via a link controller. 

The TX and RX paths of WULPUS PRO are AC-coupled to support both conventional piezoelectric transducers (PZTs) and capacitive micromachined ultrasonic transducers (CMUTs), with biasing provided by the on-board LT3463 DC-DC converter.

\subsection{Transmit path}
The WULPUS PRO system leverages the programmable pulse generator (PPG) peripheral subsystem of the MSP430 MCU to generate rectangular excitation pulses with configurable frequency (0.2 - 10\,MHz) and pulse count (0 - 127). To boost the pulse amplitude from 3.3\,V to 30\,V, a MOSFET gate driver (IXDD604) is employed. In contrast to the original WULPUS design \cite{frey2022wulpus}, this driver not only amplifies the excitation voltage but also provides tri-state pulser functionality by placing the output in a high-impedance state via a dedicated control pin. This configuration enables a single-point connection between the pulser output and the transmit-receive switch input, allowing the system to accommodate 16 transducer channels, 2$\times$ more than the original WULPUS probe using the same high-voltage multiplexer (HV2707). Furthermore, the 2$\times$ increase in excitation voltage ($15\;\text{V}\;\rightarrow\;30\;\text{V}$) enables deeper penetration and improves the signal-to-noise ratio (SNR) of the backscattered echoes.

\subsection{Receive path}

The receive path of the WULPUS PRO system comprises several components, as illustrated in Figure \ref{fig:system_diagram} (a). It begins with a pre-amplifier circuit that performs \review{amplification and} impedance matching between the transducer and the subsequent sampling stages. The PCB design supports multiple pre-amplifier configurations based on the OPA836 (Texas Instruments, Dallas, USA), with the specific implementation selected through design variants during the PCB assembly process. The two primary options are a high-input/low-output impedance voltage buffer (6\,dB gain, default) and a transimpedance amplifier, covering both PZT transducers and CMUTs. \review{A key design objective of the pre-amplifier and the following VGA stage is to provide gain in order to minimize degradation of the received echo signal SNR through the successive stages of filtering, optional envelope detection, and digitization.}

The pre-amplifier stage is followed by the Variable Gain Amplifier (VGA) AD8338, which provides programmable gain ranging from 0 to 74\,dB with a theoretical bandwidth of up to 15\,MHz. \review{Placing gain early in the receive chain reduces the contribution of downstream stages to the overall input-referred noise.} The gain is set through an analog input, where the control signal is generated by an RC network that approximates a linear amplification slope (for times $t \ll RC$). The resistance $R$ is programmable via a digital potentiometer, while the capacitance $C$ is fixed. The charging and discharging of the RC network are controlled by the digital I/Os of the MSP430 microcontroller, which operates in push-pull and high-impedance input mode. This configuration provides analog TGC for media attenuation up to 74\,dB, enabling consistent high-SNR sensing across the full imaging depth, which was not achievable with the previous design\cite{frey2022wulpus}. A fixed-gain mode is also supported by pre-charging the RC network and maintaining the charge during the acquisition.

The VGA output can be routed directly through the active low-pass filter (LPF) to the MSP430 ADC input or directed first to the LT5507 envelope detector (Analog Devices, Wilmington, USA). Enabling or bypassing the envelope-detection path is configurable at runtime.
The novel envelope extractor module extends WULPUS PRO to support high-frequency transducers (up to 8\,MHz according to SPICE simulation) whose RF signals exceed the MSP430 ADC’s Nyquist limit. It also enables bandwidth reduction, which can decrease power consumption or increase frame rate in applications where phase information is not required.

\subsection{Host interface and Wireless Link}

The host controller is connected to WULPUS PRO through a 4-wire SPI bus (8 MHz clock) and control signals host-ready and data-ready, indicating readiness of the host MCU to start the SPI transaction and availability of the ultrasound data for the readout.
 
For low pulse-repetition frequency (PRF) operation (50\,Hz), the system was evaluated using the nRF52DK development kit featuring an nRF52832 MCU (Nordic Semiconductor, Trondheim, Norway) running a modified WULPUS firmware that provides a BLE bridge (300\,kbps) to a host PC. For the new high-PRF operation mode (300\,Hz), WULPUS PRO was tested with the ESP32-C6-DevKitC-1 development board (Espressif Systems, Shanghai, China), implementing WiFi connectivity to the host PC with a measured throughput of {2}\,Mbps.

\subsection{Power Supplies}

\begin{figure}[tb]
  \centering
  \includegraphics[width=1\columnwidth, clip, trim={0 0 0 0 cm}]{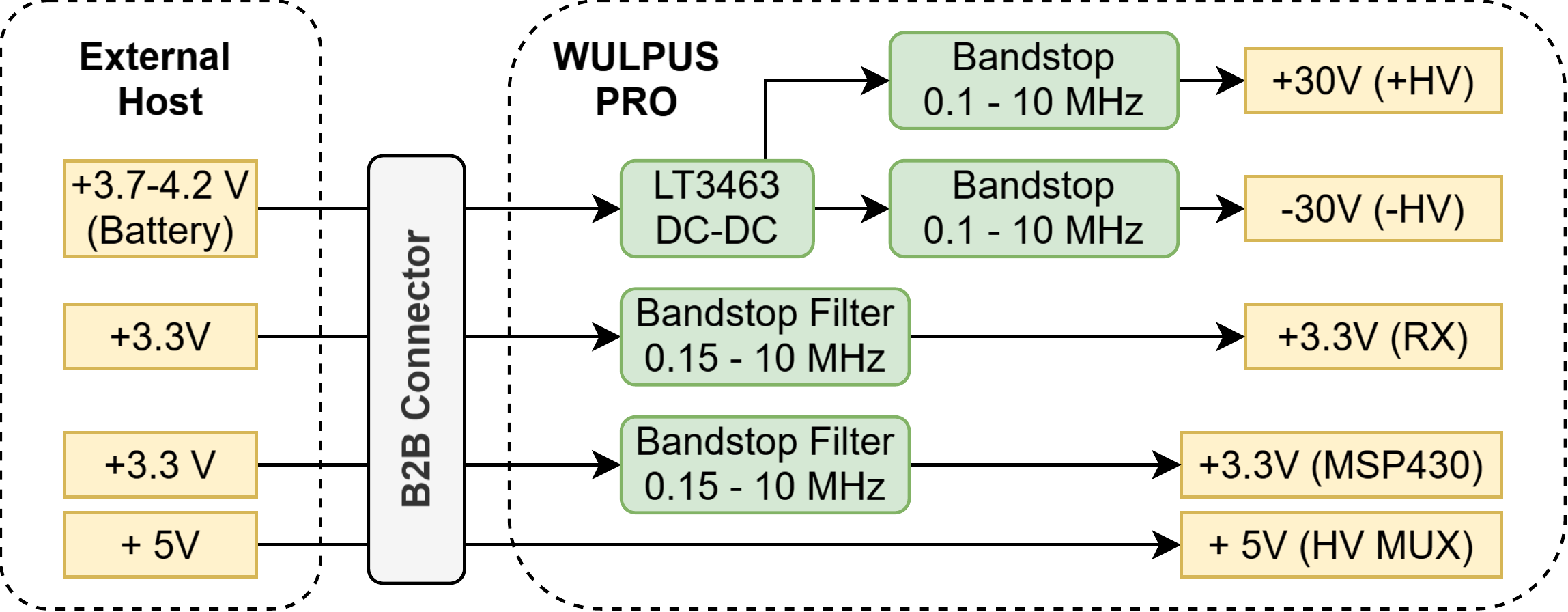}
  \caption{\review{WULPUS PRO power tree. Power is supplied from the external host through board-to-board (B2B) connector, providing battery, 3.3\,V, and 5\,V rails. The battery rail feeds the LT3463 DC-DC converter to generate the ±30\,V high-voltage supplies, while bandstop filters attenuate conducted noise within the receive bandwidth.}}
  \label{fig:power_tree}
\end{figure}

\review{The power-supply architecture of} \textsc{WULPUS PRO} \review{is intentionally kept simple to facilitate integration with a variety of host platforms and power-management schemes, while being optimized for battery-powered operation. Supply rails that are commonly available in host systems are therefore interfaced directly to avoid unnecessary conversion stages and their associated losses, whereas the more complex high-voltage rails are generated locally on the platform. As shown in Figure~\ref{fig:power_tree}, the host provides three regulated rails and one battery rail: a 3.3\,V rail for the MSP430 microcontroller and digital logic, a 5\,V rail for the high-voltage multiplexer, a low-noise 3.3\,V rail for the analog front-end, and a battery rail (3.7\,V nominal) for the LT3463 (Analog Devices, Wilmington, USA) DC-DC converter, which generates the} $\pm 30$\,V \review{high-voltage biasing and excitation domains. All critical supply rails incorporate passive bandstop filters selected to attenuate conducted noise within the receive bandwidth and reduce coupling between the power, digital, and analog domains.}

\subsection{Transducers}

The system is compatible with conventional piezoelectric transducers with frequencies up to 8\,MHz. We tested WULPUS PRO with 2.25\,MHz (LA-2.25-32, Vermon, Tours, France) and 5\,MHz (LA-5-32, Vermon, Tours, France) linear arrays featuring 32 channels. Both arrays have compact dimensions (23$\times$13$\times$10\,mm), an element pitch of 0.6\,mm, and fractional bandwidths of 89\% (2.25\,MHz) and {72}\% (5.0\,MHz), respectively. 

In addition to piezoelectric transducers, we evaluated the compatibility of WULPUS PRO with polyCMUTs. PolyCMUTs are an emerging technology for producing ultra-thin, lightweight, and cost-effective ultrasound devices~\cite{Angerer2025polyCMUT}. Fabricated through photolithographic polymer micromachining and thin-film metal deposition, they can be manufactured within 1-2 days, making them well-suited for rapid and cost-effective prototyping of wearable applications.

\begin{figure}[!b]
  \centering
  \includegraphics[width=0.98\columnwidth, clip, trim={0 0 0 0 cm}]{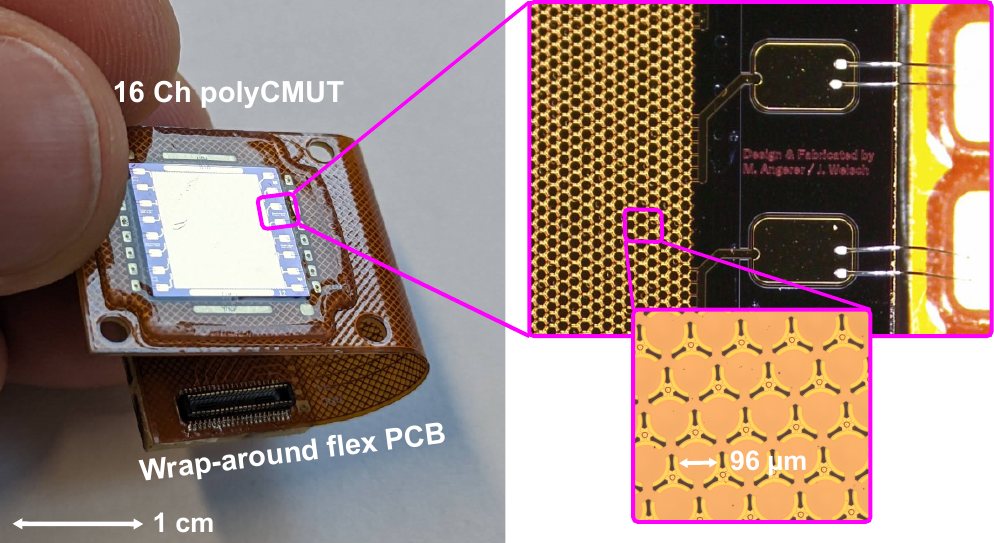}
  \caption{16 channel polyCMUT acoustic frontend with custom designed wrap-around flex PCB.}
  \label{fig:polycmut}
\end{figure}

To achieve a compact and lightweight form factor, 16-channel polyCMUT linear arrays were integrated onto custom-designed flexible PCBs that can be wrapped around the main board, resulting in a total weight of less than 2g. The polyCMUT-based acoustic frontend is shown in Figure~\ref{fig:polycmut}, including magnified views of two channels and the microscopic arrangement of miniature drum structures used for ultrasound transmission and reception. The integration of the frontend into the WULPUS PRO system is illustrated in Figure~\ref{fig:system_diagram}(c). The array features an element pitch of 0.64~mm, and the transducers operate at a center frequency of 3~MHz with a $-$6~dB fractional bandwidth of 119\% in TX \cite{Lu2025Button}. 

To seamlessly support both conventional piezoelectric transducers and CMUTs, WULPUS PRO offers versatile DC biasing options. The PCB incorporates solder jumpers that allow either indirect biasing (potential on ground electrode) up to $-$30 V, direct biasing (potential on signal electrode) up to +30 V via a bias tee, or the simultaneous use of both options. The AC coupling of the TX and RX paths to the multiplexer\footnote{ \review{The HV2707 multiplexer supports analog signal voltages up to ±100 V while requiring only a +5 V bias supply and no high-voltage supply}} channels, and consequently to the transducer outputs, protects the low-voltage electronics from the applied bias voltages. 

\subsection{Programming Model/User Interface}

\textsc{WULPUS PRO} supports configuration of the key data-acquisition parameters through a graphical interface organized into four categories: excitation, receive path, HV MUX, and advanced settings.

\review{The excitation settings configure the programmable pulse-generator frequency and pulse count, with a fixed 50\% duty cycle.} The receive-path settings define the ADC sampling frequency, PGA gain, VGA parameters, and the optional activation of the envelope extractor.

The HV MUX configuration defines the transmit-receive channel mapping. The probe may transmit on all channels while receiving on a single one, or use arbitrary subsets for both TX and RX. Up to 16 configurations can be stored in memory and executed sequentially, enabling advanced acquisition schemes such as synthetic aperture imaging.

Finally, the advanced settings control the timing of peripheral modules within the ultrasound microcontroller, including the PPG, ADC, internal bias generator, and HV MUX switching to receive mode. Further implementation details are available in the \textsc{WULPUS PRO} repository \cite{wulpus_pro_repo}.

\subsection{Tested Operation Modes}
\label{sect:implemented_modes}

To demonstrate the flexibility of the probe, we implemented a set of software-defined operating modes and characterized each by measuring their power consumption. The modes are summarized in Table \ref{table:operating_modes}, \review{and acquisitions with all three modes are described in detail in Section \ref{sect:functional_evaluation}.}

Mode I utilizes all 16 channels for \review{TX} (with one receive channel) and operates in raw RF-sampling mode at frame rates ranging from 1 to 50\,Hz, employing BLE for data transfer.
Mode II is similar to Mode I but utilizes the analog envelope extractor, enabling operation with transducers whose center frequencies exceed the ADC's Nyquist limit.
Mode III implements 16-channel synthetic aperture imaging \review{with plane-wave TX and synthetic-aperture beamforming on receive (Section \ref{sect:b-mode_imaging_pzt})} and uses Wi-Fi for data transmission. \review{The PRF is set to 300\,Hz for Mode III. Since one receive channel is acquired per TX in the 16-channel synthetic aperture sequence, the corresponding effective B-mode image rate is $PRF/16=\qty{18.75}{Hz}$. For all modes, the acquisition length is fixed to 400 samples. } 

\begingroup
\setlength{\tabcolsep}{4.7pt}

\begin{table*}[th]
\centering
\caption{Demonstrated operating modes of the WULPUS PRO system.}
\label{table:operating_modes}
\resizebox{\textwidth}{!}{%
\begin{tabular}{|c|c|c|c|c|c|c|c|c|c|}
\hline
\textbf{Mode} & 
\textbf{\begin{tabular}[c]{@{}c@{}}N receive\\ Channels\end{tabular}} & 
\textbf{\begin{tabular}[c]{@{}c@{}}PRF\\ {[}Hz{]}\end{tabular}} & 
\textbf{\begin{tabular}[c]{@{}c@{}}Transducer\\ Freq. {[}MHz{]}\end{tabular}} &
\textbf{\begin{tabular}[c]{@{}c@{}}Sampling\\ Freq. {[}MHz{]}\end{tabular}} & 
\textbf{\begin{tabular}[c]{@{}c@{}}Wireless\\ Link\end{tabular}} & 
\textbf{\begin{tabular}[c]{@{}c@{}}Capture\\ Mode\end{tabular}} & 
\textbf{TGC} & 
\textbf{\begin{tabular}[c]{@{}c@{}}Core Power\\ {[}mW{]}\end{tabular}} &
\textbf{\begin{tabular}[c]{@{}c@{}}Wireless Link\\ Power {[}mW{]}\end{tabular}} \\
\hline

\textbf{Mode I (A-mode)} & 
1 & 
50 & 
2.25 & 
8 & 
BLE & 
Raw RF & 
On & 
35 &
12 \\
\hline

\textbf{Mode II (Analog Envelope)} & 
1 & 
50 & 
2.25/5.0 & 
8 & 
BLE & 
Envelope & 
On & 
38 &
12 \\
\hline

\textbf{Mode III (2D B-mode)} & 
16 & 
300\review{$^{*}$} & 
2.25 & 
8 & 
Wi-Fi & 
Raw RF & 
On & 
58 &
314 \\
\hline

\end{tabular}
}

\par\vspace{2pt}
{\footnotesize \review{* single-channel pulse repetition frequency, equivalent to 18.75 Hz effective B-mode image rate.}}
\end{table*}

\endgroup

\section{{Hardware} Characterization}
\label{sect:electronics_characterization}

\subsection{Power Consumption}
\begin{figure}[tb]
  \centering
  \includegraphics[width=1\columnwidth, clip, trim={0 0 0 0 cm}]{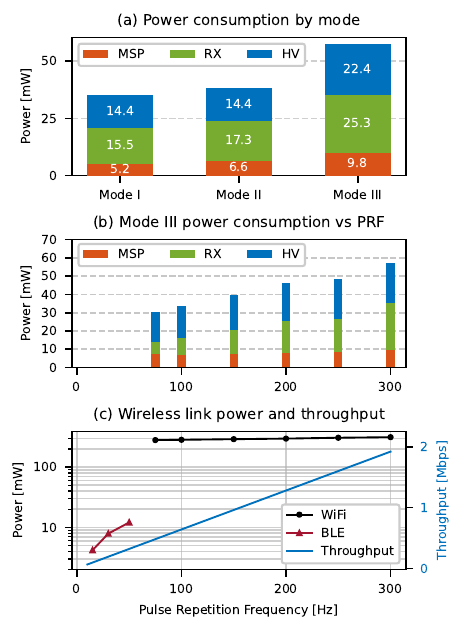}
  \caption{WULPUS PRO power consumption breakdown for all operating modes (a) (Table~\ref{table:operating_modes}), detailed power measurements for Mode III (b), and for wireless links (c) as a  function of PRF.}
  \label{fig:power_consumption_throughput}
\end{figure}

To characterize the power consumption of WULPUS PRO in each operating mode, we used a WULPUS PRO evaluation board (Fig. \ref{fig:system_diagram} (b)) and a Keysight N6705B DC power analyzer (Keysight Technologies, Santa Rosa, USA) to separately source and measure each individual power domain (MSP, HV, RX; see Fig.~\ref{fig:power_tree}). For each acquisition, the current trace of each domain was averaged over the measurement window and multiplied by its supply voltage. The power consumption of the wireless link was measured using a Nordic Power Profiler Kit II (Nordic Semiconductor, Trondheim, Norway). The results for all modes are summarized in Table~\ref{table:operating_modes} and Figure~\ref{fig:power_consumption_throughput} (a).
For Mode III, we also measured the power consumption of all power domains and the BLE/Wi-Fi modules across a PRF range of 10 to 300\,Hz, revealing which subsystems dominate power consumption and how their contributions increase with the frame rate. Figures~\ref{fig:power_consumption_throughput} (b) and (c) summarize these extended measurements.

\subsection{Receive Bandwidth and SNR Characterization}
\label{sec:receive_bandwidth}

\begin{figure}[tb]
  \centering
  \includegraphics[width=1\columnwidth, clip, trim={0 0 0 0 cm}]{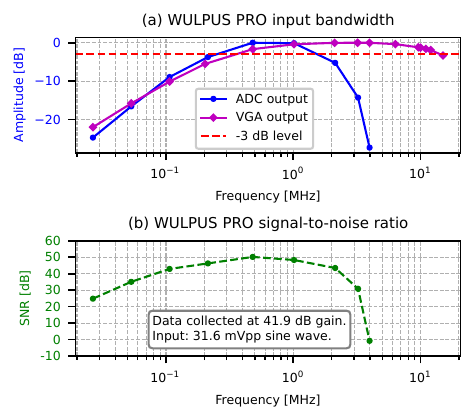}
  \caption{Receive bandwidth of the WULPUS PRO system (a) at the ADC output (blue) and VGA output (magenta), and end-to-end SNR measurements (b).}
  \label{fig:receive_bandwidth}
\end{figure}

The bandwidth and signal-to-noise ratio (SNR) of WULPUS PRO were characterized in receive-only RF mode (based on Mode I from Table~\ref{table:operating_modes}), \review{using the full receive chain including the pre-amplifier, VGA, and LPF, while} bypassing the envelope detection stage. A sinusoidal test signal generated by a Keysight 33600A waveform generator (Keysight Technologies, Santa Rosa, USA)  was injected into the transducer connector and measured both with a high-impedance \qty{10}{\mega\ohm}, \qty{9}{\pico\farad}  Sensepeek SQ500 probe on a Keysight HD304MSO oscilloscope at the VGA output and by the ADC of the MSP430 microcontroller, allowing isolated analysis of the analog amplification chain and the MSP430 ADC front-end. The sine-wave frequency was swept from 26\,kHz to 16\,MHz (for the MSP up to 4\,MHz) with a fixed amplitude of {31.6}\,mV\textsubscript{pp}. The total receive gain of the RX path was configured at {41.9}\,dB, consisting of 6\,dB from the preamplifier{, 34\,dB fixed gain from the VGA and 1.9\, dB gain from MSP430 PGA}.


For each MSP430 acquisition, the received signal was analyzed using Welch’s power spectral density (PSD) method. The signal for each frequency point was divided into overlapping 320-sample segments and windowed. The one-sided spectral density was computed for each windowed segment and averaged to obtain a smooth estimate. The maximum PSD value for each frequency point was extracted, and the combined spectrum was normalized such that the highest peak across the entire sweep corresponded to 0\,dB. For each oscilloscope acquisition, the raw data were filtered with a narrow band-pass filter around the input frequency, and the peak-to-peak sine-wave amplitudes were extracted and normalized as above. For both methods, the -3 dB\,bandwidth was then determined relative to the normalized maxima.

The signal-to-noise ratio (SNR) was derived only from the MSP430 acquisitions using the same PSD by integrating the signal and noise power. \review{The noise power was integrated over the entire one-sided Nyquist bandwidth of the sampled signal, i.e. from DC to $f_s/2$, providing a conservative worst-case estimate.} To avoid overestimating noise energy, the fundamental component and its first four harmonics, together with neighboring frequency bins, were excluded from the noise calculation. The SNR was computed as

\begin{equation}
\mathrm{SNR}_{\mathrm{dB}} =
10 \log_{10}\!\left(\frac{P_{\mathrm{signal}}}{P_{\mathrm{noise}}}\right).
\label{eq:snr_psd}
\end{equation}

\noindent where \( P_\text{signal} \) and \( P_\text{noise} \) represent the integrated powers of the signal and noise, respectively. 

Figures \ref{fig:receive_bandwidth} (a) and (b) summarize the measurements, demonstrating an end-to-end WULPUS PRO bandwidth of {1.4\,MHz (0.2\,MHz - 1.6\,MHz)} with an SNR of approximately {45\,dB} in the passband, and an amplification-path-only bandwidth of {14\,MHz (0.3\,MHz - 14.3\,MHz)}. \review{Although the -3\,dB bandwidth of the end-to-end system is limited to 1.6\,MHz by the MSP430 ADC front-end, the SNR-based characterization in Fig.~\ref{fig:receive_bandwidth}(b) shows that the signal remains above {30\,dB} up to approximately {3\,MHz}. This indicates that the practically usable receive range extends beyond the nominal -3 dB\,limit. The origin of this limitation and its implications are discussed in Section~\ref{sect:discussion}.}

\subsection{SINAD Characterization}
\label{sec:dynamic_range}

\begin{figure}[tb]
  \centering
  \includegraphics[width=1\columnwidth, clip, trim={0 0 0 0 cm}]{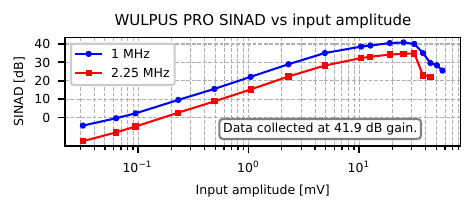}
  \caption{\review{End-to-end SINAD measurements of the WULPUS PRO system for a 1\,MHz and 2.25\,MHz sine wave over a range of input voltage amplitudes}.}
  \label{fig:receive_sinad}
\end{figure}

\review{To characterize the receive-chain noise floor and distortion-limited regime, the signal-to-noise-and-distortion ratio (SINAD) was measured as a function of input amplitude. The measurement used the same receive-chain configuration as in Section~\ref{sec:receive_bandwidth}, with the envelope detector bypassed and the raw ADC samples used for spectral analysis. The input sine-wave frequency was fixed to either \qty{1}{MHz} or \qty{2.25}{MHz}, the latter corresponding to the center frequency of the transducer used in this work. The waveform-generator output was attenuated by \qty{30}{dB} before the WULPUS PRO input.}

\review{For each input amplitude, the signal power was estimated from the fundamental component, while the remaining spectral power over the Nyquist bandwidth, excluding DC, was assigned to noise and distortion. SINAD was calculated as}

\begin{equation}
\mathrm{SINAD}_{\mathrm{dB}} =
10 \log_{10}\left(\frac{P_{\mathrm{signal}}}{P_{\mathrm{noise}} + P_{\mathrm{distortion}}}\right).
\label{eq}
\end{equation}

\review{The resulting curves are shown in Fig.~\ref{fig:receive_sinad}. SINAD increases with input amplitude in the noise-limited region, reaches a maximum of approximately \qty{41}{dB} at \qty{1}{MHz} and \qty{36}{dB} at \qty{2.25}{MHz}, and decreases at larger amplitudes due to distortion.}
\review{Peak SINAD values are reached at a {25.3\,mV input voltage amplitude for 1\,MHz and 31.6\,mV for 2.25\,MHz frequencies.}}
\section{Functional Evaluation}
\label{sect:functional_evaluation}

\subsection{Time Gain Compensation}
\label{sect:tgc_validation}

We validated the time-gain compensation (TGC) functionality using the CIRS 040GSE tissue-mimicking phantom (Sun Nuclear, USA). The 2.25\,MHz transducer was positioned above the vertical scatterer group within the high-attenuation zone. Every four neighboring  elements of the original transducer (2.25\,MHz, 32 ch.)  were connected in parallel, forming eight groups that were wired to the \textsc{WULPUS PRO} input channels. The probe operated in Mode I (Table~\ref{table:operating_modes}), employing all channels during transmission, while the central group (corresponding to a single \textsc{WULPUS PRO} channel) was selected for reception.

\begin{figure}[tb]
  \centering
  \includegraphics[width=1\columnwidth, clip, trim={0 0 0 0 cm}]{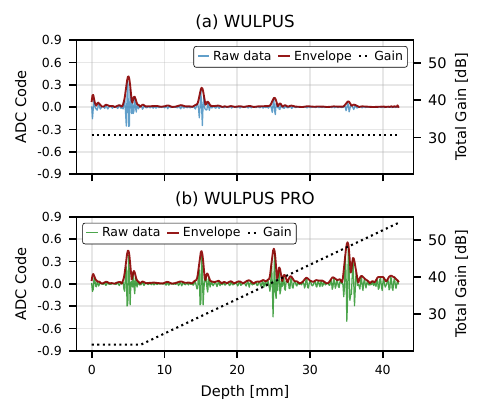}
  \caption{TGC performance of WULPUS PRO (b) compared to the fixed-gain performance of WULPUS (a). Raw data are band-pass filtered from 1.3 to 3.6\,MHz, and the envelope is extracted using a Hilbert transform.}
  \label{fig:tgc_comparison}
\end{figure}

First, the original WULPUS probe was tested with a fixed total gain of 30\,dB (PGA + LNA). Subsequently, the WULPUS PRO probe operated in Mode I (Table~\ref{table:operating_modes}) with a TGC profile achieving a 40\,dB gain at the midpoint of the acquisition (see Fig.~\ref{fig:tgc_comparison}). The acoustic scene was identical in both measurements, as the transducer position remained unchanged and only the probe cables were reconnected.

The results in Fig.~\ref{fig:tgc_comparison} show that, for the original WULPUS, the amplitude of the fourth scatterer is heavily attenuated relative to the first one. In contrast, WULPUS PRO fully compensates for this attenuation, increasing the gain from approximately 21 to 53\,dB with a linear slope. Moreover, WULPUS PRO requires roughly 9\,dB less receive gain to achieve the same echo amplitude, enabled by its 2× higher excitation amplitude (30\,V) and a cleaner receive path compared to the original WULPUS design.

\subsection{Analog Envelope Extraction}
\label{sect:analog_envelope_validation}

We also validated the analog envelope extractor using the same setup as for the TGC validation described in Sect.~\ref{sect:tgc_validation}. To demonstrate its effectiveness, both 2.25\,MHz and 5\,MHz transducers were employed, operating in Mode II (see Tab.~\ref{table:operating_modes}). For each transducer, the raw RF data was acquired first, followed by a second acquisition in which the output of the envelope detector was digitized.

\begin{figure}[tb]
  \centering
  \includegraphics[width=1\columnwidth, clip, trim={0 0 0 0 cm}]{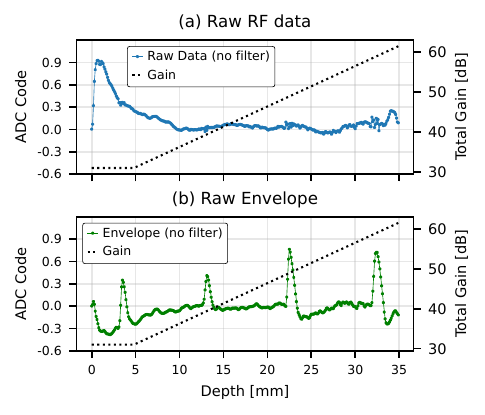}
  \caption{Analog envelope extractor (b) vs. RF data (a) for the 5.0\,MHz transducer.}
  \label{fig:env_5_0_MHz}
\end{figure}

\begin{figure}[tb]
  \centering
  \includegraphics[width=1\columnwidth, clip, trim={0 0 0 0 cm}]{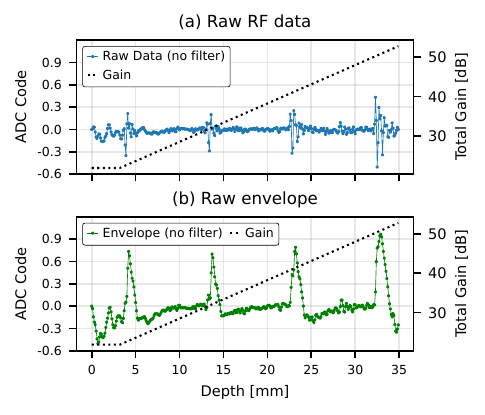}
  \caption{Analog envelope extractor (b) vs. RF data (a) for the 2.25\,MHz transducer.}
  \label{fig:env_2_25_MHz}
\end{figure}

The experimental results are presented in Fig.~\ref{fig:env_2_25_MHz} and Fig.~\ref{fig:env_5_0_MHz}. For the 5\,MHz transducer (Fig.~\ref{fig:env_5_0_MHz}), no echoes are visible in the raw RF waveform, as the signal components lie above the Nyquist frequency. In contrast, the output of the envelope extractor exhibits distinct peaks corresponding to the scatterers. For the 2.25\,MHz transducer, the raw RF data, as expected, show visible echo signals, while the envelope extractor output also reveals well-defined signatures of the vertical scatterers.

\subsection{B-mode Imaging}
\label{sect:b-mode_imaging_pzt}

\begin{figure}[tb]
  \centering
  \includegraphics[width=1\columnwidth, clip, trim={0 0 0 0 cm}]{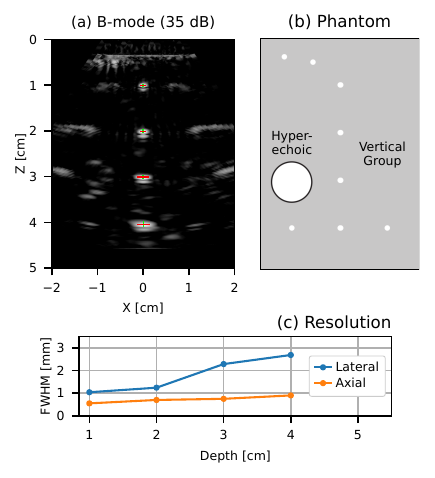}
  \caption{Vertical scatterer group of the CIRS phantom: B-mode image with a 2.25 MHz piezoelectric transducer (a), phantom diagram (b), and extracted axial and lateral resolution (c)}
  \label{fig:vertical_scatters}
\end{figure}

\begin{figure}[tb]
  \centering
  \includegraphics[width=1\columnwidth, clip, trim={0 0 0 0 cm}]{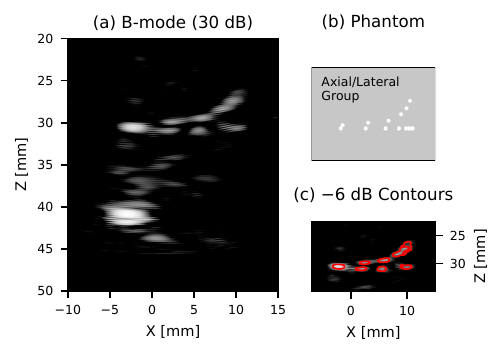}
  \caption{Axial–lateral scatterer group of the CIRS phantom: B-mode image with a 2.25 MHz piezoelectric transducer (a), phantom diagram (b), and $-$6 dB contours (c).}
  \label{fig:axial_lateral_scatters}
\end{figure}

To demonstrate the B-mode imaging capabilities of the 16-channel configuration, the 2.25\,MHz transducer was used. To map
the 32 available transducer channels to the 16 system channels,
we paired adjacent elements, resulting in an effective pitch of
1.2 mm. The vertical scatterer group (Fig.~\ref{fig:vertical_scatters} (b)) and axial–lateral scatterer group (Fig.~\ref{fig:axial_lateral_scatters} (b)) of the CIRS 040GSE phantom were imaged using synthetic aperture method with a single plane wave transmission. For image reconstruction, the open-source PyBF \cite{pybf} framework was modified to support a Delay-and-Sum (DAS) beamformer with coherence factor weighting \cite{hollman1999coherence} and \review{dynamic receive} apodization \cite{tomov2004compact}.

Following image reconstruction, the axial and lateral resolutions were estimated from the vertical scatterer group using the full width at half maximum (FWHM) approach. The reconstructed image of the vertical scatterer group is shown in Fig.~\ref{fig:vertical_scatters}. At a depth of 3\,cm, the system achieves approximately 0.7\,mm axial resolution and 2.3\,mm lateral resolution in the center of the beam.

The axial–lateral scatterer group, located at a 3\,cm depth, was additionally analyzed to confirm these findings by evaluating scatterer separability through visualization of the $-$6\,dB contours (equivalent to the full width at half maximum, FWHM) extracted around the reflection maxima. The reconstructed image of the axial–lateral scatterer group is shown in Fig.~\ref{fig:axial_lateral_scatters}. The measurements indicate slightly better performance than for vertical scatterer group, demonstrating axial separation for scatterers spaced as closely as 0.5\,mm and lateral separation up to 2\,mm.

\section{Integration with polyCMUTs}
\label{sect:integ_polycmuts}

After verifying the performance of the WULPUS PRO platform with the commercially available piezoelectric transducers, we proceeded to evaluate the polyCMUT acoustic frontend. The frontend was operated using a -30\,V indirect bias generated by the WULPUS PRO electronics, applied to the ground electrodes of the array. The excitation signal (0–30\,V, AC-coupled) was applied to the top electrodes, corresponding to the individual channels. To assess performance, we conducted pulse-echo measurements followed by preliminary B-mode imaging experiments. 

\subsection{Pulse Echo Testing}
\label{sect:polycmuts_pulse}

The polyCMUT frontend introduced earlier in Figure \ref{fig:polycmut} was connected to a WULPUS PRO test board (Figure \ref{fig:system_diagram} (b) and (c)) and dipped in a dish filled with non-conductive isopropyl alcohol to minimize the risk of electrical shorts. A 3\,cm thick aluminum block was positioned at a distance of about 1\,cm from the transducer to serve as an acoustic reflector. 

\begin{figure}[tb]
  \centering
  \includegraphics[width=1\columnwidth, clip, trim={0 0 0 0 cm}]{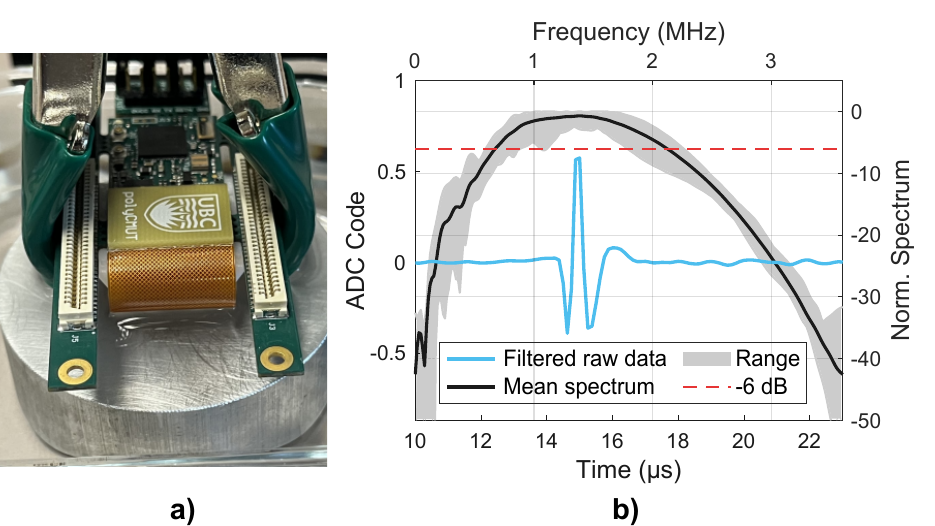}
  \caption{Pulse-Echo testing with the polyCMUT acoustic frontend: (a) Measurement setup with Aluminum reflector. (b) Single-channel pulse echo, resulting spectrum and variations over all 16 channel.}
  \label{fig:polycmut_txrx}
\end{figure}

Figure~\ref{fig:polycmut_txrx} shows the test setup, the measured echo signal and the resulting mean spectrum and total range from the 16 measured channels. The spectral analysis of the received pulses yielded a center frequency of 1.46\,MHz, and a mean $-$6 dB fractional bandwidth of 96\%. The discrepancy between the transducer’s nominal center frequency (3\,MHz \cite{Lu2025Button}) and the measured center frequency arises from the system’s receive bandwidth limitations, as illustrated in Figure~\ref{fig:receive_bandwidth}.

\subsection{B-mode Imaging}
\label{sect:polycmuts_bmode}

\begin{figure}[tb]
  \centering
  \includegraphics[width=1\columnwidth, clip, trim={0 0 0 0 cm}]{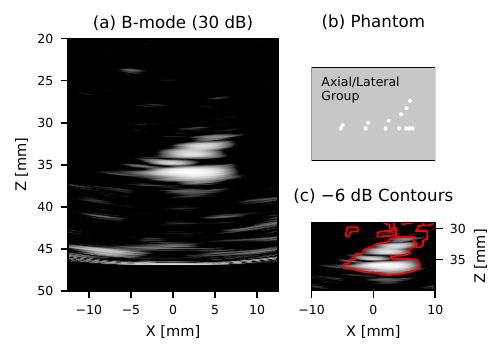}
  \caption{Axial–lateral scatterer group of the CIRS phantom captured with the polyCMUT acoustic frontend: reconstructed B-mode image (a), phantom diagram (b), and –6\,dB contours (c).}
  \label{fig:axial_lateral_scatters_polycmut}
\end{figure}

To evaluate the imaging capabilities of the \mbox{polyCMUT} acoustic frontend, the axial–lateral scatterer group of the CIRS phantom was again scanned using the acquisition techniques and image reconstruction approach described in Section~\ref{sect:b-mode_imaging_pzt}. The resulting B-mode image is shown in Figure~\ref{fig:axial_lateral_scatters_polycmut}. Compared to the piezoelectric transducers shown in Figure~\ref{fig:axial_lateral_scatters}, the \mbox{polyCMUT} acoustic frontend achieves comparable axial resolution but lower lateral resolution. This is a result of the smaller overall aperture (10.2\,mm for the \mbox{polyCMUT} versus 19.2\,mm for the LA-2.25-32) and the reduced element pitch, which were selected to realize a more compact system footprint.

\section{Comparison to the State-of-the-Art}
\label{sect:soa_comparison}

\begingroup
\setlength{\tabcolsep}{4.7pt}

\begin{table*}[th]
\centering
\caption{Performance comparison to state-of-the-art wearable US research platforms.}
\label{table:competitors}
\resizebox{\textwidth}{!}{%
\begin{tabular}{|c|l|c|c|c|l|c|c|c|c|c|c|}
\hline
\multirow{4}{*}{\textbf{Platform}} & 
\multicolumn{1}{c|}{\multirow{4}{*}{\textbf{\begin{tabular}[c]{@{}c@{}}Size {[}mm{]}\\weight {[}g{]}\end{tabular}}}} & 
\multirow{4}{*}{\textbf{\begin{tabular}[c]{@{}c@{}}\# TX \\ chan.\end{tabular}}} & 
\multirow{4}{*}{\textbf{\begin{tabular}[c]{@{}c@{}}Transmit \\ Path\end{tabular}}} &
\multirow{4}{*}{\textbf{\begin{tabular}[c]{@{}c@{}}\# RX \\ chan.\end{tabular}}} &
\multicolumn{1}{c|}{\multirow{4}{*}{\textbf{\begin{tabular}[c]{@{}c@{}}Receive \\ Path\end{tabular}}}} & 
\multirow{4}{*}{\textbf{\begin{tabular}[c]{@{}c@{}}Data \\ link\end{tabular}}} & 
\multirow{4}{*}{\textbf{\begin{tabular}[c]{@{}c@{}}Transfer\\ rate $R$\\ {[}Mbps{]}\end{tabular}}} & 
\multirow{4}{*}{\textbf{\begin{tabular}[c]{@{}c@{}}PRF \\ {[}Hz{]}\end{tabular}}} & 
\multirow{4}{*}{\textbf{\begin{tabular}[c]{@{}c@{}}$\overline{P}$\\ {[}mW{]}\end{tabular}}} & 
\multirow{4}{*}{\textbf{\begin{tabular}[c]{@{}c@{}}$P_{ch}$\\[2pt] $\left[\dfrac{\textrm{mW}}{\textrm{chan.}}\right]$\end{tabular}}} & 
\multicolumn{1}{c|}{\multirow{4}{*}{\textbf{Modes}}} \\
 & \multicolumn{1}{c|}{} &  &  & \multicolumn{1}{c|}{} &  &  &  &  & & & \multicolumn{1}{c|}{} \\ 
  & \multicolumn{1}{c|}{} &  &  & \multicolumn{1}{c|}{} &  &  &  &  & & & \multicolumn{1}{c|}{} \\ 
   & \multicolumn{1}{c|}{} &  &  & \multicolumn{1}{c|}{} &  &  &  &  & & & \multicolumn{1}{c|}{} \\ 
\hline

\multirow{4}{*}{\textbf{\begin{tabular}[c]{@{}c@{}}WULPUS PRO\\ (this work)\end{tabular}}} & 
\multirow{4}{*}{\begin{tabular}[c]{@{}l@{}}\textit{L:} 39 \\ \textit{W:} 21\\ \textit{H:} 6\\ \textit{m:} 5 g\end{tabular}} & 
\multirow{4}{*}{16} & 
\multirow{4}{*}{\begin{tabular}[c]{@{}c@{}}30\,V, unipolar,\\ configurable, \\ PZTs $\&$ CMUTs \end{tabular}} & 
\multirow{4}{*}{1} &
\multirow{4}{*}{\begin{tabular}[c]{@{}l@{}}8\,Msps 12-bit,\\ LNA (6\,dB), \\ VGA (60\,dB)\\TGC\end{tabular}} & 

\multirow{2}{*}{BLE} &
\multirow{2}{*}{0.3} & 
\multirow{2}{*}{\begin{tabular}[c]{@{}c@{}}50 \\ (1 ch.)\end{tabular}} & 
\multirow{2}{*}{35} & 
\multirow{2}{*}{35} &
\multirow{4}{*}{\begin{tabular}[c]{@{}c@{}}A, \\ B\\ M \\ Env.\end{tabular}}
\\ 
& & & & & & 
 & & & & &
\\
\cline{7-11}

 &  &  &  &  & & 
\multirow{2}{*}{WiFi} &
\multirow{2}{*}{2} 
 & \multirow{2}{*}{\begin{tabular}[c]{@{}c@{}}300 \\ (1 ch.)\end{tabular}} & 
 \multirow{2}{*}{58} & 
 \multirow{2}{*}{58} &
 \\ 
 & & & & & & 
 & & & & &
\\
\hline

\begin{tabular}[c]{@{}c@{}}WULPUS\\ \cite{frey2022wulpus}\end{tabular} & \begin{tabular}[c]{@{}l@{}}\textit{L:} 46\\ \textit{W:} 25\\ \textit{H:} 13\\ \textit{m:} 13 g\end{tabular} & 8 & \begin{tabular}[c]{@{}c@{}}15\,V, unipolar,\\ configurable \\ PZTs\end{tabular} &
1 &\begin{tabular}[c]{@{}l@{}}8\,Msps 12-bit,\\ LNA (10\,dB), \\ PGA (30.8\,dB)\end{tabular} & \begin{tabular}[c]{@{}c@{}} BLE\end{tabular} & 0.3 & \begin{tabular}[c]{@{}c@{}}50 \\ (1 ch.)\end{tabular} & 13 & 13 & \begin{tabular}[c]{@{}c@{}}A, \\ M\end{tabular} \\ \hline

\begin{tabular}[c]{@{}c@{}}USop\\ \cite{lin2024fully}\end{tabular} & \begin{tabular}[c]{@{}l@{}}$^{\mathrm{*}}$\textit{L:} 90\\ \textit{W:} 36\\ \textit{H:} 5\\ \textit{m:} -\end{tabular} & 32 & \begin{tabular}[c]{@{}c@{}}70-100\,V, unipolar,\\ fixed freq. and\\ pulse,  PZTs\end{tabular} & 
1 & \begin{tabular}[c]{@{}l@{}}12\,Msps 12-bit,\\ 44\,dB gain \\ (fixed),\\ band pass filter\end{tabular} & \begin{tabular}[c]{@{}c@{}} WiFi\end{tabular} & 3.4 & - & 300 & 300 & \begin{tabular}[c]{@{}c@{}}A, \\ B,\\ M\end{tabular} \\ \hline

\multirow{4}{*}{\begin{tabular}[c]{@{}c@{}}TinyProbe\\ \cite{vostrikov2024tinyprobe}, \cite{hirschi2025high}\end{tabular}} & 
\multirow{4}{*}{\begin{tabular}[c]{@{}l@{}}\textit{L:} 57 \\ \textit{W:} 35\\ \textit{H:} 20\\ \textit{m:} 35 g\end{tabular}} & 
\multirow{4}{*}{32} & 
\multirow{4}{*}{\begin{tabular}[c]{@{}c@{}}$\pm$32\,V, bipolar,\\ configurable,\\ TX beamforming \\ PZTs\end{tabular}} & 
\multirow{4}{*}{32} &
\multirow{4}{*}{\begin{tabular}[c]{@{}l@{}}30\,Msps, 10-bit\\ LNA (21\,dB), \\ PGA (27\,dB), \\TGC\end{tabular}} & 
\multicolumn{1}{c|}{\multirow{4}{*}{\begin{tabular}[c]{@{}c@{}} WiFi\end{tabular}}} & 
\multirow{2}{*}{21.6} & 
\multirow{2}{*}{\begin{tabular}[c]{@{}c@{}}175 \\ (32 ch.)\end{tabular}} & 
\multirow{2}{*}{430} & 
\multirow{2}{*}{13.4} &
\multirow{4}{*}{\begin{tabular}[c]{@{}c@{}}A, \\ B, \\ M,\\ Doppler\end{tabular}} 
\\ 
& & & & & & \multicolumn{1}{c|}{} & & & & &\\
\cline{8-11}
 &  &  &  &  & & \multicolumn{1}{c|}{} & \multirow{2}{*}{11.6} 
 & \multirow{2}{*}{\begin{tabular}[c]{@{}c@{}}1400 \\ (2 ch.)\end{tabular}} & 
 \multirow{2}{*}{830} & 
 \multirow{2}{*}{26} &
 \\ 
 & & & & & & \multicolumn{1}{c|}{} & & & & & \\
\hline

\begin{tabular}[c]{@{}c@{}} \cite{speicher2025wearable}\end{tabular} & \begin{tabular}[c]{@{}l@{}}\textit{L:} 184\\ \textit{W:} 123\\ \textit{H:} 33\\\textit{m:} 610 g\end{tabular} & 128 & \begin{tabular}[c]{@{}c@{}}$\pm$ 100\,V, bipolar,\\ configurable,\\ TX beamforming, \\ PZTs\end{tabular} &
32$^{\dag}$ &\begin{tabular}[c]{@{}l@{}}50\,Msps \\ (12-bit)\end{tabular} & \begin{tabular}[c]{@{}c@{}}WiFi\end{tabular} & 21.6 & \begin{tabular}[c]{@{}c@{}}40 \\ (32 ch.)\end{tabular} & 10100 & 316 & \begin{tabular}[c]{@{}c@{}}A, \\ B, \\ M\end{tabular} \\ \hline

\begin{tabular}[c]{@{}c@{}} PuLsE \\ \cite{giordano2025pulse}\end{tabular} & \begin{tabular}[c]{@{}l@{}}\textit{L:} -\\ \textit{W:} -\\ \textit{H:} -\\\textit{m:} -\end{tabular} & 1 & \begin{tabular}[c]{@{}c@{}}$\pm$ 15\,V, bipolar,\\ configurable,\\ PZTs\end{tabular} &
1 &\begin{tabular}[c]{@{}l@{}}4\,Msps 12-bit, \\ Envelope \\detector\end{tabular} & \begin{tabular}[c]{@{}c@{}}BLE\end{tabular} & - & \begin{tabular}[c]{@{}c@{}}25 \\ (1 ch.)\end{tabular} & 5.8 & 5.8 & \begin{tabular}[c]{@{}c@{}}A, M \\ (Env.)\end{tabular} \\ \hline

\multicolumn{12}{l}{\rule{0pt}{3ex} $\overline{P}$: average core power (excluding wireless link). $P_{ch}$: power per receive channel.}\\

\multicolumn{12}{l}{\rule{0pt}{3ex}$^{\mathrm{*}}$Estimated from the drawings in the Supplementary Information of \cite{lin2024fully}, \rule{0pt}{3ex}$^{\dag}$32 RX electronics channels are multiplexed to a 128-channel array. 
}\\

\end{tabular}
}
\end{table*}

\endgroup

Table~\ref{table:competitors} compares the performance of the WULPUS PRO platform with other ultrasound research systems, focusing on battery-powered wearable solutions, their feature sets, supported transducer technologies, and core power consumption. Wireless link power is excluded from the analysis, as its optimization lies outside the scope of this work and represents a separate, application-dependent engineering challenge.

At one end of the design spectrum are wearable ultrasound systems tailored for specific applications and ultra-low-power operation. An example is PulsE~\cite{giordano2025pulse}, developed for heart-rate monitoring, which combines a 10\,MHz single-channel transducer with analog envelope-detection electronics and consumes only 5.8\,mW at a 25\,Hz PRF. WULPUS PRO differentiates itself by providing access to raw data from multiple channels at higher acquisition rates, enabling a broader range of research use cases at the expense of higher baseline power consumption.

Compared to the UsoP platform, WULPUS PRO offers significantly broader runtime configurability, including programmable excitation parameters, flexible receive configurations (channel selection and gain settings), and adjustable acquisition timing. Although UsoP provides a 40\% higher sampling rate (12\,Msps vs. 8\,Msps) and demonstrates operation with a 6\,MHz transducer, its per-channel power consumption is 5–8× higher, significantly limiting battery life.

Building on the foundation of the original system, WULPUS PRO introduces substantial feature-set improvements over its predecessor, the original WULPUS platform \cite{frey2022wulpus}. In particular, the doubled excitation amplitude, doubled channel count, and added TGC support enable deeper B-mode imaging by providing stronger insonification, a wider effective receive aperture, and compensation for depth-dependent tissue attenuation. Although power consumption increases relative to the original WULPUS design, this is primarily due to the higher excitation energy (2× voltage and 2× channel count), while the overall budget remains below 50\,mW per receive channel at PRFs up to 250\,Hz (Fig.~\ref{fig:power_consumption_throughput}(a)). Consequently, WULPUS PRO supports all applications previously demonstrated with WULPUS hardware, such as cardiorespiratory monitoring and gesture recognition, while also enabling new use cases thanks to its enhanced specifications.

The TinyProbe system~\cite{vostrikov2024tinyprobe} represents a high-end wearable ultrasound platform, offering best-in-class per-channel power consumption and the most comprehensive research feature set and mode support. WULPUS PRO complements TinyProbe by targeting applications  where a lower channel count (16 vs. 32) and lower-frequency transducers (below 8\,MHz) are sufficient, but compact size, low weight, and low total power consumption ($<$100\,mW) are prioritized.

Across all systems discussed, WULPUS PRO appears to be the first fully long-term wearable platform, supporting both piezoelectric and CMUT transducer technologies, enabled by its AC-coupled TX/RX paths and flexible biasing options.

\section{Discussion}
\label{sect:discussion}

The most important constraint of modern wearable ultrasound systems remains the limited power budget, which \review{restricts} the feasible channel count, operating frequencies, and \review{functionality} of the acquisition system. \review{Although power requirements vary across wearable applications depending on the imaging modality and acquisition parameters, minimizing the average power consumption benefits all these systems by extending operating time and reducing battery weight.} The core electronics of WULPUS PRO consume only 35 mW in A-mode (Mode I, Table~\ref{table:operating_modes}). When paired with a low-power BLE wireless link (Table~\ref{table:operating_modes}), the system enables continuous raw data streaming at a 50\,Hz PRF for more than 24 hours from a 300\,mAh Li-Po battery. For high-PRF applications, WULPUS PRO was coupled with an industry-standard Wi-Fi 6 module, supporting raw data streaming up to 300\,Hz PRF (Mode III, Table~\ref{table:operating_modes}). In this configuration, the WULPUS PRO electronics consume only 58\,mW, enabling multi-hour continuous operation. Wireless-link efficiency and battery life could be further improved by adopting modern Wi-Fi 6 solutions with target wake time support \cite{hirschi2025high} or next-generation BLE 5.4 SoCs (e.g., nRF54L15). However, these considerations are left for future work, as the present study focuses on the ultrasound acquisition frontend.

Compared to the previous design \cite{frey2022wulpus}, in WULPUS PRO, the transmit path was reworked to increase the excitation amplitude to \qty{30}{V_{pp}} and to enable more TX channels, thereby producing a stronger acoustic response and contributing to a higher receive echo amplitude. In addition, the RX path architecture (amplification chain, power supply filters, and PCB layout) was redesigned to provide more features (TGC, envelope detector) and a cleaner signal chain with low noise and high bandwidth. Validation measurements demonstrate an amplification-path-only bandwidth of {14 MHz} and an end-to-end \review{-3 dB} bandwidth of 1.4\,MHz with an SNR of approximately {45 dB} in the passband at {41.9\,dB} total gain. This bandwidth limitation (1.4\,MHz) originates from the sigma–delta ADC of the MSP430 microcontroller, inherited from the original WULPUS design, which employs a seventh-order CIC decimation filter introducing approximately –6.4\,dB attenuation at 2 MHz (see Formula 11, p.~572 of the MSP430 Family Datasheet). \review{However, the -3\,dB bandwidth alone does not fully describe the practically usable operating range. When the usable bandwidth is instead assessed from the measured SNR, the signal remains above a practical measurement threshold ($\geq \qty{30}{dB}$) up to {3}\,MHz. This is consistent with the results in Fig.~\ref{fig:axial_lateral_scatters}, which show successful synthetic-aperture B-mode operation with a 2.25\,MHz transducer despite operation beyond the nominal -3\,dB limit.}

As the ADC constitutes the primary bandwidth bottleneck of the system, future revisions plan to replace MSP430 with a higher-bandwidth data-conversion solution such as e.g. STM32L412 (ST Microelectronics, Switzerland). Like the STM32L496 used in PuLsE \cite{giordano2025pulse}, this MCU belongs to the same ultra-low-power family and features two peripheral ADC modules capable of operating in interleaved mode at 12.3\,Msps (10-bit). Preliminary measurements confirm a bandwidth up to the Nyquist frequency of $\approx6$\,MHz, making it a strong candidate for front-end performance improvement.

An alternative approach to ease the high-bandwidth requirements of modern ultrasonic transducers is to reduce signal bandwidth in the analog domain by performing envelope extraction prior to digitization, as introduced in \cite{giordano2025pulse}. WULPUS PRO implements this concept using the LTC5507 RF power detector for amplitude demodulation of echo signals, demonstrated in phantom measurements with 2.25\,MHz (Fig.~\ref{fig:env_2_25_MHz}) and 5\,MHz (Fig.~\ref{fig:env_5_0_MHz}) transducers. While higher-frequency transducers were not available for experimental validation, SPICE simulations confirm the operation of the envelope extractor for transducers with central frequencies up to 8 MHz, which is also within the bandwidth of the amplification chain (see Fig. \ref{fig:receive_bandwidth} (a)). According to the datasheet, the LTC5507 supports an envelope bandwidth of up to 1.5\,MHz, which is sufficient to track the amplitude variations of typical 2–8\,MHz echo pulses after removal of the RF carrier. 
\review{This allows the MSP430 ADC to sample the demodulated envelope rather than the original RF waveform, thereby making more effective use of the available ADC bandwidth. However, this mode is not suitable for synthetic-aperture B-mode imaging, since envelope detection removes the phase coherence required for coherent beamforming. The envelope path is therefore primarily intended for amplitude-based acquisition scenarios, such as A-mode measurements.} Transmitting only the envelope also reduces data throughput and wireless power consumption. With an active power consumption of less than 2\,mW and a deep power-down mode, the LTC5507-based envelope path of WULPUS PRO is well-suited for applications that rely solely on amplitude information. 

In addition to supporting higher-frequency transducers via envelope extraction, WULPUS PRO enhances performance at greater depths through the implementation of time-gain compensation (TGC). The lightweight TGC design, based on the AD8338 VGA, effectively compensates for depth-dependent attenuation of up to 70\,dB, as demonstrated in Fig.~\ref{fig:tgc_comparison} and Fig.~\ref{fig:env_5_0_MHz} using the vertical scatterer group of the tissue-mimicking phantom CIRS 040GSE (high-attenuation zone). In WULPUS PRO, careful component selection and use of the VGA power-down mode minimize the additional power overhead, making TGC practical within a wearable power budget ($\leq$50\,mW for Modes I and II; see Table~\ref{table:operating_modes}).

All the new capabilities of the WULPUS PRO system enabled, for the first time, plane-wave B-mode imaging based on an ultra-low-power wearable ultrasound platform, achieving a frame rate of 18 FPS using 16-channel synthetic-aperture acquisition at a 300\,Hz PRF. We adapted an open-source beamforming framework \cite{pybf} to implement a DAS-CF beamformer for low-channel-count operation, as CF weighting helps emphasize spatially coherent echoes and suppress less coherent clutter and sidelobe artifacts, which can be more pronounced with a limited 16-channel aperture. B-mode imaging was demonstrated  on a tissue-mimicking phantom, achieving an axial resolution of $\approx$ 0.7\,mm and a lateral resolution of $\approx$ 2.3\,mm at the center of the field of view with a 2.25\,MHz transducer (Fig.~\ref{fig:vertical_scatters} and Fig.~\ref{fig:axial_lateral_scatters}). Although the results are limited by the end-to-end system bandwidth, comparison with TinyProbe~\cite{vostrikov2024tinyprobe} (using DAS beamforming) shows a comparable axial resolution (0.8\,mm for TinyProbe) and a modest degradation in lateral resolution (1.6\,mm for TinyProbe), indicating a high signal quality achievable with WULPUS PRO despite its substantially lower power budget.

The lower lateral resolution of the B-mode image obtained with the polyCMUTs (Figure~\ref{fig:axial_lateral_scatters_polycmut}), compared to the piezoelectric transducer array (Fig. ~\ref{fig:axial_lateral_scatters}) can be primarily attributed to the smaller effective aperture, as the polyCMUT acoustic frontend exhibits only half the element pitch. The associated reduction in active area per channel also leads to a lower sensitivity, increasing the noise in the reconstructed image. Hence, achieving good lateral resolution at high SNR requires a large array aperture, which conflicts with a compact system design. From an imaging perspective, the pitch of the resulting elements quickly exceeds one wavelength in water. However, since the system is not intended for high-end imaging with a high channel count, the associated grating lobes may represent an acceptable trade-off. With their small form factor, low weight (2\,g), potential skin conformability, and low-cost production capacity, polyCMUTs represent a highly promising acoustic frontend technology for this system. Future work will focus on designing and producing custom arrays specifically tuned to the operating frequency, size requirements, and imaging constraints of WULPUS PRO to fully leverage the system’s capabilities \review{including future validation with skin-conformal transducer implementations.} \review{Additionally, future work will evaluate the long-term reliability under both direct and indirect DC bias conditions.}

Finally, WULPUS PRO features a compact size ($39\times21\times6$\,mm, 5\,g) and modular architecture that allows pairing with arbitrary wireless or wired data-transmission modules and seamless integration into other research hardware. Decoupling the ultrasound front end from the communication interface maximizes flexibility, enabling application-specific optimization of the data link without modifying the core ultrasound electronics.

\section{Conclusion}
\label{sect:conclusion}

This work presents WULPUS PRO, a next-generation research acquisition platform that advances the capabilities of ultra-low-power wearable ultrasound probes. With a compact size of $39\times21\times6\;\mathrm{mm}$ and low weight of 5\,g, the system integrates 16 time-multiplexed channels, a two-state configurable pulser (up to 30\,V), a receive front end with up to 70\,dB total gain (LNA + VGA), time-gain compensation support, an optional analog envelope extraction module, and an 8-Msps ADC providing a {14\,MHz} amplification path and 1.4-MHz end-to-end bandwidth.

We present several features that demonstrate improved B-mode imaging performance in ultra-low-power wearable ultrasound devices.
These include runtime-configurable TGC, enhanced power supply filtering, a low-noise amplification chain, and an optimized PCB layout, which collectively provide a clean RX path with a mid-band SNR of {45\,dB} and effective compensation for depth-dependent tissue attenuation. Together, these elements enable 16-channel B-mode synthetic-aperture imaging of deeper targets on an ultra-low-power platform ($<$50\,mW for 50\,Hz PRF). \review{Phantom results demonstrate sub-millimeter axial resolution and millimeter-scale lateral resolution, indicating that WULPUS PRO can achieve useful B-mode image quality within the constraints of a compact, low-channel-count wearable architecture.}

To extend support for higher-frequency transducers, \mbox{WULPUS PRO} integrates an ultra-low-power envelope detector circuit, \review{which demodulates the RF signal before digitization and reduces the required sampling bandwidth. This supports} transducers up to approximately 8\,MHz (with 1.5\,MHz envelope bandwidth) for applications that require only amplitude information. The module consumes less than 2\,mW in active mode and substantially reduces data throughput, lowering wireless-link utilization and prolonging battery life.

Despite its significantly expanded feature set, WULPUS PRO remains highly power efficient, consuming $<$50\,mW at a 50\,Hz PRF and $<$60\,mW even at a high PRF of 300\,Hz. This level of efficiency enables at least all-day BLE operation and multi-hour WiFi operation when the system is interfaced with the corresponding external wireless modules, all powered from a compact 300\,mAh LiPo cell.

The system offers versatile DC biasing options, supporting a range of ultrasound transducer technologies. Full compatibility is demonstrated with both piezoelectric transducers and polymer-based CMUTs, with the latter achieving a high fractional transmit bandwidth of 94\%. With its integrated and validated set of features, WULPUS PRO establishes a configurable platform for small-scale, lightweight, and skin-conformal applications for long-term continuous ultrasound monitoring.

\section*{Statement of Contributions}

Conceptualization, [S.V., A.C.]; methodology, [S.V.]; hardware design and implementation, [S.V., F.V., A.C.]; PolyCMUT design and fabrication, [M.A., J.W., J.L.]; software and firmware development, [S.V., C.H.]; investigation, [S.V., C.H., J.L., F.V.]; validation, [S.V., F.V., C.H., J.L.]; formal analysis, [S.V., C.H., J.L.]; data curation, [S.V., C.H., J.L.]; visualization and graphics, [S.V., M.A.]; writing—original draft preparation, [S.V., A.C., M.A., F.V., C.H.]; writing—review and editing, all authors; supervision, [L.B., A.C., R.R., E.C.]; project administration, [S.V., A.C., M.A.]; funding acquisition, [A.C., M.A., L.B.]. All authors have read and agreed to the published version of the manuscript.

\section*{Acknowledgment}

The authors would like to thank Sebastian Frey (ETH Z{\"u}rich)
for schematic review and technical comments, Alfonso Blanco Fontao
(ETH Z{\"u}rich) for coordinating PCB fabrication, and Ciara Giles
Doran for a preliminary investigation of the AD8338 VGA conducted
as part of a semester project.

\bibliographystyle{IEEEtran}
\bibliography{bibliography}

\end{document}